\documentclass[12pt,draft,aps,prb,floats]{revtex4}
\usepackage[dvips]{epsfig}
\usepackage{graphicx}  
\begin{document}

\title{\bf Short-time dynamics of polypeptides}

\author{Everaldo Arashiro}
\affiliation{John v. Neumann Institute for Computing (NIC),
             Forschungszentrum J{\"u}lich, 52425 J{\"u}lich, Germany}
\email{e.arashiro@fz-juelich.de}

\author{J. R. Drugowich de Fel\'{\i}cio}
\affiliation{Departamento de F\'{\i}sica e Matem\'{a}tica,
             FFCLRP Universidade de S\~{a}o Paulo, Avenida Bandeirantes 3900,
             CEP 14040-901,  Ribeir\~{a}o Preto, S\~ao Paulo, Brazil}
\email{drugo@usp.br}

\author{Ulrich H.E. Hansmann}
\affiliation{John v. Neumann Institute for Computing (NIC),
             Forschungszentrum J{\"u}lich, 52425 J{\"u}lich, Germany}
\affiliation{Department of Physics, Michigan Technological University,
         Houghton, MI 49931-1291, USA}
\email{hansmann@mtu.edu, u.hansmann@fz-juelich.de}

\begin{abstract}
 The authors study the short-time dynamics of  helix-forming polypeptide chains
using an all-atom representation of the molecules and an implicit
solvation model to approximate the interaction with the surrounding
solvent. The results confirm earlier observations that the
helix-coil transition in proteins can be described by a set of
critical exponents. The high statistics of the simulations allows
the authors to determine the exponents values with increased
precision and support universality of the helix-coil transition in
homopolymers and (helical) proteins.
\end{abstract}
\date{\today}
\maketitle

\section{INTRODUCTION}

While energy landscape theory and funnel concept
\cite{Bryngelson87,Onuchic97} are powerful tool for describing the
general characteristics of folding,  many details remain to be
solved. One example is the nature of the transitions between
different thermodynamic states that are observed during folding of
proteins. Their nature and  role in the folding process are  active
areas of research as many diseases are associated with spontaneous
or induced misfolding of proteins. In principle, these transitions
can be probed {\it in silico}. However, the energy landscape of
proteins is  characterized by a rough energy landscape with a huge
number of local minima separated by high energy barriers.
Considerable effort has been put into the development of sampling
techniques that overcome the resulting slow convergence of computer
experiments \cite{H03a,Review1}.   In generalized-ensemble
techniques \cite{Review2} such as parallel tempering \cite{PT}, the
broad-histogram method \cite{Oliveira} or multicanonical
sampling\cite{MU} , simulations are performed in an {\it artificial}
ensemble where the convergence to the equilibrium distribution is
fast. Thermodynamic quantities are then calculated through
re-weighting to the canonical (i.e., physical) ensemble. In
Ref.~\onlinecite{AHD06} we applied a complimentary approach to
protein science that  leaves the ``physical" dynamics unchanged. The
underlying idea is to extract information on certain equilibrium
properties from the critical relaxation of regular canonical Monte
Carlo  simulations (i.e., a non-equilibrium behavior of the system).

Here, we revisit this approach and study with much improved
statistics the helix-coil transition in simple polypeptides. Since
it allows us to compare our results with earlier work,
 we study in the present paper the nonequilibrium evolution process of
$(Ala)_N$ chains ($N=10, 20, 30 $ and $40$) in short-time Monte
Carlo (MC) simulations. Our investigation is later extended towards
the 34-residue  human parathyroid fragment  PTH(1-34). We evaluate
and compare for both systems the exponents that characterize the
observed helix-coil transition.

\section{Methods}

Janssen {\it et al.}  \cite{Janssen1,Huse}
have shown that at criticality universality and scaling  are already
present early in  the evolution of systems.
A system, characterized by an order parameter $m$, decays at
a critical temperature $T = T_c$ from an initially ordered state $m_0=1$
 (after a microscopic time $t_{mic}$) according to
 a power-law \cite{Stauffer92,Stauffer93,Sch96} $m(t) \sim t^{-\beta / \nu z}$.
  This relation is obtained from the  more general scaling form:
\begin{equation}
 m(t,\tau ,N)=
 b^{-\beta /\nu } m (b^{-z}t, b^{1/\nu }\tau,b^{-1}N), \label{magk}
\end{equation}
where $b$ is an arbitrary chain size scaling factor
and $\tau $ is the reduced temperature, $\tau =(T-T_{c})/T_{c}$.
Differentiating $\ln m(t,\tau)$ with respect to the temperature $T$ at
$T=T_{c}$,
\begin{equation}
\left. {\displaystyle{\partial \ln m(t,\tau,L) \over \partial\tau}}
\right|_{\tau=0}   \sim  t^{1/\nu z}  \, ,               \label{derivate}
\end{equation}
leads  to a scaling relation that allows one to extract the exponent
\cite{Li96} $1/\nu z$ . Combining the two  scaling relations allows
one further to eliminate the dynamic exponent $z$ and to calculate
the estimates of the critical exponent $\beta$. Additional
information on the system can be obtained from the scaling
\cite{Review} of the second cumulant
\begin{equation}
U_2(t,L)=\frac{m^2(t,L}{(m(t,L))^2} \sim t^{d/z}.
\label{second_moment}
\end{equation}
We use  in the present paper Eq.~\ref{second_moment} in combination
with Eq.~\ref{derivate} to obtain estimates of the exponent
$d\nu$.

Another dynamic critical exponent was proposed
by Majumdar et al. \cite{Majumdar1996}. The authors studied
 global persistence probability P(t) which  is   defined as
\begin{equation}
P(t)=1-\sum_{t^{\prime }=1}^{t}\rho (t^{\prime }),
\end{equation}
where $\rho (t^{\prime })$ is the fraction of the samples where the
order parameter $m$ has changed its sign for the first time at the
instant $t^{\prime }$.
At criticality, P(t) is expected to decay algebraically as
\begin{equation}
P(t) \sim t^{-\theta _{g}} \label{Persistence}
\end{equation}
where $\theta _{g}$\ is the global persistence exponent. The study
of the persistence behavior have attracted enormous interest,
playing an important role in the study of systems far from
equilibrium\cite{Majumdar2003,Schulke1997,Hinrichsen1998,Zheng2002,daSilva2003,
HenriqueeArashiro,Henrique2006} .

The question arises whether the above scaling  laws apply also
to transitions in proteins and other biological
molecules for that - as finite systems - the concept of a phase transition
is not defined. For instance, the formation of $\alpha$-helices
in proteins   resembles crystallization. The nature of this helix-coil transition
is a long-standing topic of research \cite{Scheraga1} and has been studied
with various methods including renormalization group  techniques \cite{Scheraga2}.
A few years ago additional evidence was presented  by one of us \cite{HO98c,AH98,AH00b}
 that this helix-coil transition is for polyalanine
(a homopolymer where  in principle one can define the limes of an infinite
chain) a true thermodynamic phase transition.
 Because of these previous results we
decided to focus on helix-formation  in order to study the
scaling behavior of short-time dynamics in proteins.

We define our order parameter as  the number of helical residues
 $q_H = \max\left((2<n_H(T)>/(N-2) - 1), 0\right)$.  Here we define a residue
as helical  if its backbone dihedral angles $(\phi,\psi)$ take
values in the range \cite{OH95b} $(-70^{\circ}\pm
30^{\circ},-37^{\circ}\pm30^{\circ})$ and the residue is hydrogen
bonded. The normalization factor $N -2$ is chosen instead of $N$,
the number of residues,  because the flexible terminal residues are
usually not part of an $\alpha$-helix. Our definition ensures that $
0\le q_H \le 1$ and $q_h (T_c) = 0$.

Our short-time MC simulations of the helix-coil transition
are based on a detailed, all-atom representation of proteins.
The interaction between the atoms  is
described by a standard force field, ECEPP/2 ({\bf E}mpirical
{\bf C}onformational {\bf E}nergy {\bf P}rogram for {\bf P}eptides,
version 2) \cite{EC} as implemented in the program package SMMP
({\bf S}imple {\bf M}olecular {\bf M}echanics for {\bf P}roteins)
 \cite{SMMP}:
\begin{eqnarray}
E_{ECEPP/2} & = & E_{C} + E_{LJ} + E_{HB} + E_{tor},\\
E_{C}  & = & \sum_{(i,j)} \frac{332q_i q_j}{\varepsilon r_{ij}},\\
E_{LJ} & = & \sum_{(i,j)} \left( \frac{A_{ij}}{r^{12}_{ij}}
                                - \frac{B_{ij}}{r^6_{ij}} \right),\\
E_{HB}  & = & \sum_{(i,j)} \left( \frac{C_{ij}}{r^{12}_{ij}}
                                - \frac{D_{ij}}{r^{10}_{ij}} \right),\\
E_{tor}& = & \sum_l U_l \left( 1 \pm \cos (n_l \chi_l ) \right).
\label{ECEPP/2}
\end{eqnarray}
Here, $r_{ij}$ (in \AA) is the distance between the atoms $i$ and $j$, and
$\chi_l$ is the $l$-th torsion angle.
  The interactions between our polypeptides  and water are approximated by
means of an implicit water model which assumes that the solvation
(free) energy is proportional to the solvent accessible surface area:
\begin{equation}
   E_{solv}=\sum_i\sigma_i A_i,
\end{equation}
 where $E_{solv}$ is the solvation energy, $A_i$ is the
solvent accessible area of the $i$-th atom,
and $\sigma_i$ is the corresponding solvation parameter.
 We have chosen the parameter set of Ref. \onlinecite{OONS}
that is often used in conjunction with the ECEPP force field.

\section{Results and Discussion}
We start our investigation  of the short time dynamics in the
helix-coil transition  by simulating polyalanine chains of lengths
$N=10,20,30,$ and $40$.  Our results are averaged over 4000
independent runs for $N=10,20,30$ and 2000 runs for $N=40$. Errors
are estimated  by dividing these 4000 (2000) runs in bins of 100
(50) runs and calculating the fluctuation of the averages calculated
for each bin. Figure~1a displays for various temperatures the time
series of our order parameter $q_h$ as a function of Monte Carlo
time for Ala$_{10}$. Corresponding plots are displayed for
Ala$_{20}$ (Ala$_{30}$) in Fig.~1b (1c),  and for Ala$_{40}$ in
Fig.~1d.

We expect to see in such log-log plot that curves of $q_H$
corresponding to low temperatures approach a constant value, while
high-temperatures are characterized by an exponential decrease of
$n_H$. Both temperature regions are separated by the critical
temperature $T_c$ where the corresponding curve is a straight line
(in the scaling region). Our plots in Fig.~1 indicate for $N=10$ as
critical temperature
 $T^{10}_c = 315$ K,  $T^{20}_C = 415$ K for $=20$ , $T^{30}_c = 450$ K for
$N=30$ ;  and one of $T^{40}_c= 470$ K  for our longest chain, Ala$_{40}$.

These estimates of the critical temperatures  are lower than the
ones presented in Ref.~\onlinecite{PHA03}, where for $N=10$, $N=20$ and
$N=30$ the following temperatures were estimated from the position
of the maximum in the specific heat:
 $T^{10}_c = 333(2)$ K, $T^{20}_c = 430(2)$ K and $T^{30}_c = 461(2)$ K.
Hence, for these chain sizes the estimates obtained from a
short-time analysis seem to be lower by $\approx 15$ K than the ones
estimated from the peak in specific heat. This does not indicate
that one of the two sets is wrong. Even in spin systems the estimate
of the critical temperature derived from the peak in the specific
heat usually differs for small system size from the one obtained
from the change in magnetization. Only for large system size both
values converge.  However, we remark that our values are calculated
over a set of 4000 (2000 for $N=40$) {\it independent} runs while
the results of multicanonical runs  in Ref.~\onlinecite{PHA03}
contain only of order $\approx 10$ independent events.

Note that the scaling relation
leads to similar results for all chain lengths. This can be seen in Fig.~2
which displays for $T=T_c$  the time series of the order parameter as a
function of time for all four chain lengths. We focus here on the linear
range and show both the measured data points and the straight line that
best fits these data.  The data for the different chain sizes
are almost parallel to each other and therefore differ little in their slope.
From the slope of these plots we can obtain an estimate of $\beta/\nu z$.

Measuring in addition the slope for slightly higher (lower)
temperatures and calculating the numerical derivative leads in
addition to an estimate for $1/\nu z$. In Fig.~3, the power law
increase of Eq. (2) is plotted in double-log scales for the for
polyalanine chains. Estimates of the ratio  $d/z$ can be extracted
from the scaling of $U_2(t,L)$ which is shown for the four
polyalanine chains in Fig.~4.  Table 1 lists the   values of all
three ratios $\beta/\nu z$, $1/\nu z$ and $d/z$, together with the
exponents $\beta$ and $d\nu$ that can be extracted from them.

In order to
estimate the  exponent $\theta_g$ one needs to prepare the initial states
carefully to obtain precise values of the initial $q_h^0 = q_h (t=0)$. The
exponent $\theta_g$ is obtained from Eq. ~\ref{Persistence}.
 Fig.~5 shows the corresponding log-log plots of $P(t)$
for {\it Ala}$_{40}$ and various initial values of $q_h^0$. From
these curves one extracts in the limes $N \rightarrow \infty$ and
$q_h^0 \rightarrow 0$ an
estimate for the persistence exponent $\theta_g =0.73(1)$.
The corresponding values for  polyalanine chains
of length $N=10,20,30$ and $N=40$, and for the polypeptide PTH(1-34),
are also listed in table 1.

Our values for the critical exponent $\beta \approx 0.39$ in table 1
clearly excludes   $\beta = 0$, the value expected for a first order
transition. A similar statement holds for $d\nu \approx 1.4$ that
also excludes a first-order transition ($d\nu = 1$).  This is different
 from the case of polyalanine in
gas phase where the helix-coil transition is of first-order type\cite{AH98}.

The estimates listed in table 1 indicate that the
finite size effects are less than the statistical errors and
that reliable estimators for
critical exponents  can already be obtained on very small chains.
This suggests that short-time analysis can also be applied to proteins
where an extrapolation to infinite chain
length is not possible. Here, short-time analysis  opens a way to
characterize thermodynamic processes in proteins by defining
analogues to phase transitions.

We have checked this assumption for the peptide fragment PTH(1-34)
corresponding to residues 1-34 of human parathyroid
hormone \cite{Klaus,Crystal,MABFR}. This fragment is sufficient for the
biological activities of the hormone suggesting medical and
pharmaceutical applications of the peptide. The structure of PTH(1-34) has
been resolved both crystallized \cite{Crystal} and in solution \cite{MABFR}
and  is at room temperature almost completely helical.
In previous work \cite{H03} using the same energy function it was shown
that PTH(1-34) exhibits at $T=560(10)$ K a sharp transition between  a
high temperature region where disordered coil structures prevail, and a
low-temperature region that is characterized by mostly helical structures.
The nature of this helix-coil transition could not be established in
Ref.~\onlinecite{H03}.

Observing the   short-time dynamics of 500 runs for the helix-coil
transition in PTH(1-34) we obtained data for temperatures
$T=520,530,540,545,550$ and $560$ K. Fig.~6 displays  our order
parameter for various temperatures as a function of Monte Carlo
time.  From the plot we estimate that the critical temperature is
$T=545(5)$ K, a value that is again slightly smaller than the
obtained from the peak in specific heat. The scaling region is shown
for this temperature in Fig. 7, with both the data point and the
best straight-line fit through them.

Note that our estimate of the critical temperature differs slightly
from the one ($T=540$ K) reported in Ref.~ \onlinecite{AHD06}. The
slightly lower value that was obtained in this preliminary work with
much lower statistics led to critical exponents different from the
ones for polyalanine. Our new data  indicate however that the
critical exponents (listed also in table 1) are the same for the
polyalanine chains and the polypeptide PTH(1-34). This is an
interesting result as it indicates that the order of the helix-coil
transition does not depend on the side chains of the residues in a
protein. On one hand we have (poly)alanine with its extrem short
side chain, on the other PTH(1-34) where the side chains differ  for
each residue and therefore side chain ordering is much more
difficult to obtain than in a homopolymer. Hence, our new results
suggest  ``universality'' of the helix-coil transition in
homopolymers and (helical) proteins.

\section{Summary and Conclusion}
We have investigated the short-time dynamics of helix-forming polypeptides
with up to  $N=40$ residues. At their critical temperature the evolution of these
molecules can be described by scaling laws that allow on to extract various
critical exponents  that characterize the helix-coil transition in these
molecules. Our results indicate ``universality''  of the helix-coil transition in
protein-like molecules. They also demonstrate that the analysis of short-time
dynamics is a valuable tool for the investigation of transitions in proteins that
may lead to a deeper understanding of their folding mechanism.


\section{Acknowledgment}

E. Arashiro and J. R. Drugowich de Fel\'{\i}cio acknowledge support
by Capes and CNPq (Brazil), and U.H.E Hansmann by a research grant
from the National Science Foundation (CHE-0313618).

\newpage



\newpage \cleardoublepage

\noindent
{\Large Tables}\\
\begin{table}[h]
\begin{tabular}{c|c|c|c|c|c|c}
\hline
 $N$ &  $\beta/\nu z$  &  $ 1/\nu z $    & $d/z$ &   $\beta$ & $d\nu$ & $\theta_g$\\
\hline
 10  &   0.303(3)      &   0.772(7)     & 1.099(8) & 0.392(7)   & 1.42(3) & 0.402(6)\\
 20  &   0.304(3)      &   0.782(9)     & 1.100(9) & 0.389(8)   & 1.40(3) & 0.687(7)\\
 30  &   0.301(5)      &   0.779(9)     & 1.09(1)  & 0.39(1)    & 1.40(3) & 0.708(7)\\
 40  &   0.299(5)      &   0.78(1)      & 1.09(1)  & 0.38(1)    & 1.42(4)  & 0.710(8)\\
\hline
PTH  &   0.290(5)      &   0. 768(8)      & 1.09(1) & 0.38(1)   & 1.42(3)  & 0.705(6)\\
\hline
\end{tabular}
\caption{Exponents as obtained for polyalanine chains of length
         $N=10,20$ and $40$ from the scaling relations of
         Eqs.~\ref{magk}, \ref{derivate}, \ref{second_moment} and \ref{Persistence}.
         The  columns 5 and 6  list the critical exponents $\beta$ and
        $d\nu$  as calculated from these quantities. The last
         row summarize the results for the polypeptide PTH(1-34).}

\end{table}

\newpage
\noindent
{\Large Figures}\\
\begin{description}
\item [Fig.~1] Log-log plot of the time series of the helical order parameter $q_H$
               (defined in the text) as a function of Monte Carlo time for
               polyalanine chains of length (a) $N=10$, (b) $N=20$, (c) $N=30$;
               and (d) $N=40$.
\item [Fig.~2] Log-log plot of the time series of the helical order parameter $q_H$
               as a function of Monte Carlo time for polyalanine chains of length
               $N=10,20,30$ and $40$ as measured at their respective critical
                temperatures. Shown are the data in the scaling region
               and the best fit through them.
\item [Fig.~3] Log-log plot of the time series of the logarithmic derivative of
               the order parameter (Eq.\ \ref{derivate})
               as a function of Monte Carlo time  for polyalanine chains of length
               $N=10,20,30$ and $40$, and  for the polypeptide PTH(1-34).
\item [Fig.~4] Log-log plot of the time series of the second moment $U_2(t,L)$
               as a function of Monte Carlo time for polyalanine chains of length
               $N=10,20,30$ and $40$ and the polypeptide PTH(1-34) as measured
               at their respective critical temperatures. Shown are the data in the scaling region
               and the best fit through them.
\item [Fig.~5] Log-log plot of the time series of the global persistency $P(t)$
                        (defined in the text) for {\it Ala}$_{40}$ and various initial
                        order parameter values $q_h^0$.
\item [Fig.~6] Log-log plot of the time series of the helical order parameter $q_H$
               as a function of Monte Carlo time for the polypeptide PTH(1-34)
               measured at temperatures $T =520,530,540,545,550$ and $560$ K.
\item [Fig.~7] Log-log plot of the time series of the helical order parameter $q_H$
               as a function of Monte Carlo time for the polypeptide PTH(1-34)
               at the critical temperature $T_c = 545$
               in the scaling region, with the best fit also drawn through them.
\end{description}

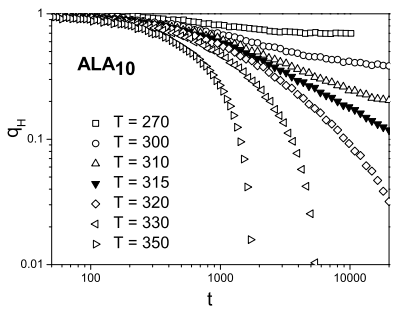
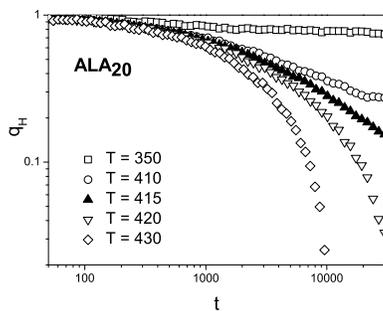
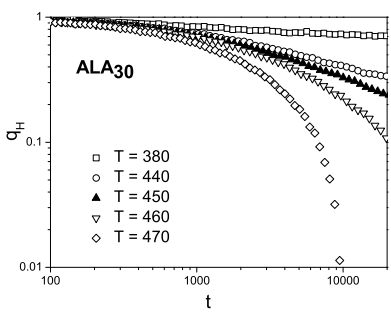
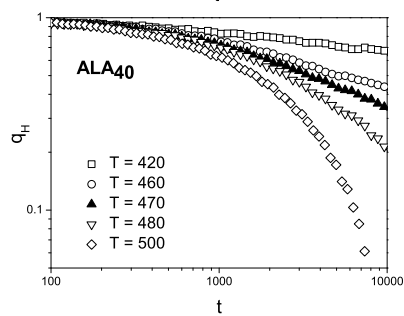
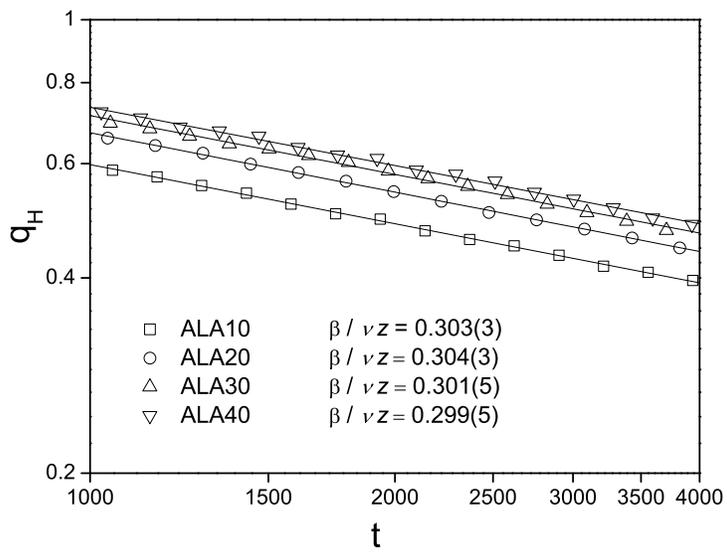
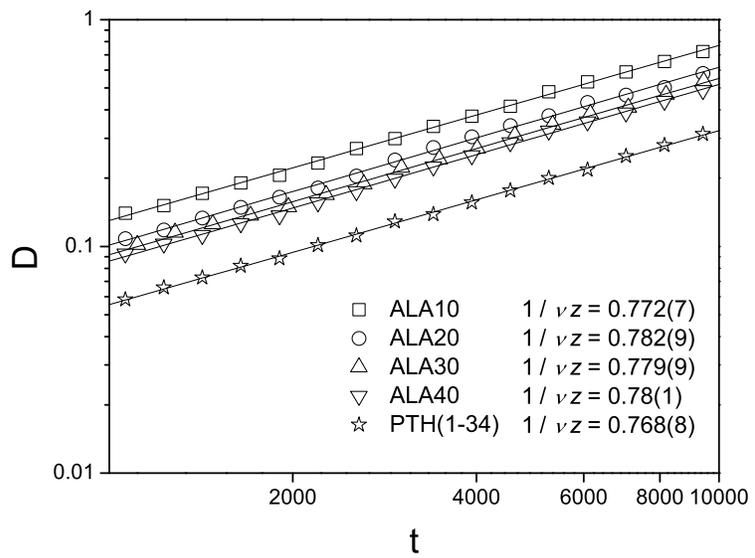
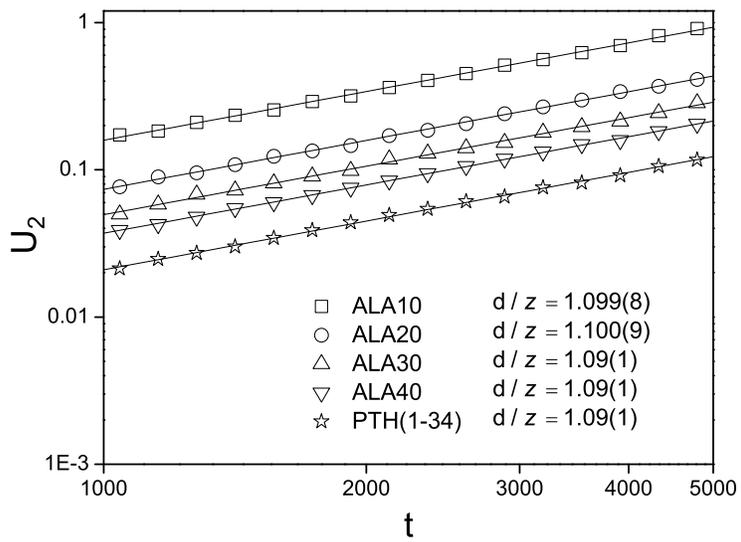
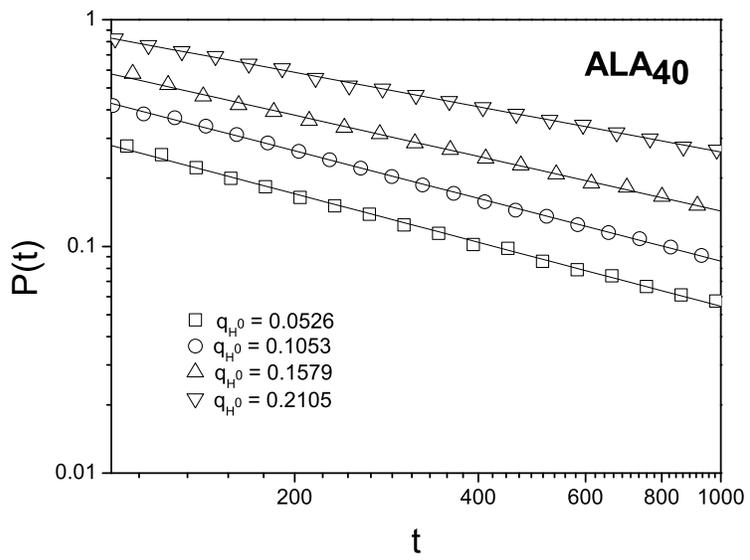
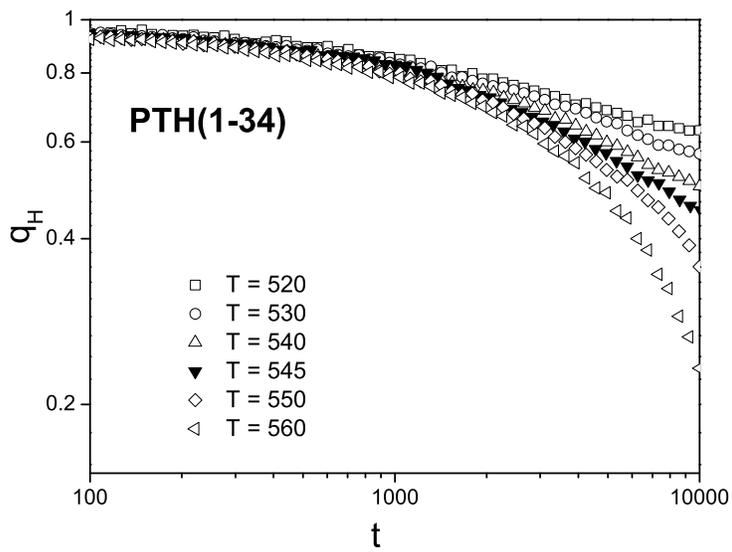
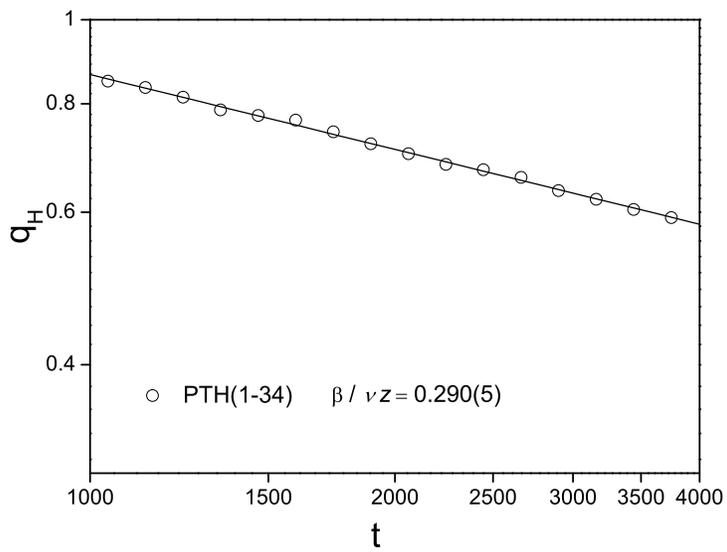
\end{document}